\documentclass[aps,pra,preprint,showpacs]{revtex4-1}
\usepackage{amsmath,graphicx,epsfig}
\begin{document}
\def\la{{\langle}}
\def\ra{{\rangle}}
\def\q{{\quad}}
\def\n{\\ \nonumber}
\def\al{{\alpha}}

\title{ Quantum measurements, stochastic networks, the uncertainty principle, and the not so strange "weak values"}
\author {D. Sokolovski}
\affiliation{Departamento de Qu\'imica-F\'isica, Universidad del Pa\' is Vasco, UPV/EHU, Leioa, Spain}
\affiliation{IKERBASQUE, Basque Foundation for Science, E-48011 Bilbao, Spain}

   \date{\today}
   \begin{abstract}
   Suppose we make a series of measurements on a chosen quantum system.
The outcomes of the measurements form a sequence of random events which occur in a particular  order. 
The system, together with a meter or meters, can be seen as following the paths of a stochastic network connecting all possible outcomes. The paths are shaped from the virtual paths of the system, and the corresponding probabilities are determined  by the measuring devices employed. If the measurements are highly accurate, the virtual paths become "real", 
and the mean values of a quantity (a functional) is directly related to the frequencies with which the paths are travelled.
If the measurements are highly inaccurate, the mean (weak) values are expressed in terms of the relative probabilities amplitudes. For pre- and post-selected systems they are bound to take arbitrary values, depending on the chosen transition.
This is a direct consequence of the uncertainty principle, which forbids one to distinguish between interfering alternatives, 
while leaving the interference between them intact.
\end{abstract}

\maketitle
\section{Introduction}
A sequence of measurements made on a quantum system leads to number of random outcomes, observed by an
experimentalist. 
In this sense, the composite system+meter(s) may be seen as following a path across a stochastic network, connecting all possible measurement results. The same can be said about a purely classical system, and the quantum nature of the experiment reveals itself in the manner, in which the transition probabilities are calculated. 
While a classical theory allows for non-invasive observations, a quantum meter plays an active role in "shaping" the network. 
In the simplest case of von Neumann  (vN) measurements \cite{vN},  the transition probabilities, as well as the observed events themselves, are determined by  the virtual paths available to the studied system, the nature of the measured quantity, and the accuracy of the meter(s).

Perhaps the most studied network is produced by  an accurate vN measurement of a projector on a state $|\psi\ra$ (preparation, or pre-selection), followed by a not necessarily accurate vN measurement of an operator $\hat{A}$, followed by an accurate measurement  of a projector on a final state $|\phi\ra$ (post-selection) \cite{Ah}. Post-selection may succeed or fail, and a sub-ensemble
is selected by collecting the statistics only in the case it is successful. If $\hat{A}$ has $N$ distinct eigenvalues, there a $N$ virtual paths connecting the two states, for which quantum mechanics provides probability amplitudes, but not probabilities, if the system is considered in isolation. If the measurement of $\hat{A}$ is accurate (ideal, strong), the meter destroys the interference between the paths, which can be now equipped with probabilities  (squared of the moduli of the corresponding amplitudes), and become "real". Each time the experiment is repeated, 
the pointer points at one of the eigenvalues of the operator $\hat{A}$, whose mean value is then calculated using the relative frequencies of these occurrences. The probability to end up in $|\phi\ra$ is not what it would be, had the meter not perturbed the system's evolution.

More controversial is the opposite limit, where the intermediate measurement is made inaccurate (weak), in order to avoid the said perturbation. The weakness of the measurement inevitably causes the meter's readings to be spread, covering the whole real axis in the limit when the perturbation is sent to zero \cite{Ah}. This gives an operational meaning to the uncertainty principle, which states that the value of $\hat{A}$ in a superposition of its eigenstates must be {\it indeterminate}. Indeed, trying to measure it without destroying the superposition would yield any value at all.

Much of the controversy resides, however, in the description of the mean of a weak meter's publication of \cite{Ah}, it gave rise a number  of intriguing concepts. These include "negative kinetic energy" \cite{NKE}, "negative number of particles" \cite{AhHARDY}, "having one particle in several places simultaneously" \cite{AhBOOK}, "photons disembodied from its polarisation" \cite{CAT}, 
"electrons with disembodied charge and mass" \cite{CAT}, and "an atom with the internal energy disembodied form the mass" \cite{CAT}, all supported by the "evidence" of weak measurements.

Recently we have shown \cite{PLA16} that the above statements amount to over-interpretations, easily dismissed once the weak values are identified with the transition amplitudes on the paths connecting the initial and final states, and are not seen as the values of the measured quantities. In this paper we will continue to look at the problem of consecutive quantum measurements, and the classical statistical ensembles produced whenever a meter, or meters, interact with the observed quantum system. 
We will pay special attention to the importance of the quantum uncertainty principle, and emphasise the role quantum interference 
plays in the loss of information about the system's past. Throughout the paper, we will use the simplest case of a two-level system as an example, and refer the reader to Refs. \cite{fst} -\cite{flast} for a more general analysis. 
% Additional work on the weak measurements 

The rest of the paper is organised as follows. In Sect. II we compare two of the best known formulations of the uncertainty principle.
In Sect. III we briefly describe the work of a von Neumann meter. In Section IV we introduce a simple classical toy model, 
to be compared with its quantum counterparts later. 
Section V returns to the simplest quantum network produced by a vN measurement of a quantity, with pre- and post-selection.
In Section VI we use our approach to describe a measurement of the difference of two physical quantities, in such a way that the values themselves remain indeterminate. Section VII sets the rules for combining virtual paths. Accurate (strong), and inaccurate (weak) limits of the measurements 
are analysed in Sections  VIII and IX, respectively. In Sect. X the uncertainty principle is used to explain the properties of the weak values.  Common misconceptions related with the issue are discussed in Sect. XI. Section XII contains our conclusions.
%%%%%%%%%%%%%%%%%%%%%%%%%%%%%%%%%%%%%%%%%%%%%%%%%%
\section {Quantum uncertainty principle(s)}
There are several formulations of the uncertainty principle (UP) covering various aspects of the problem (for a recent review see \cite{UNC}). 
Perhaps best known is the one relating the respective standard deviations, $\sigma_A$ and $\sigma_B$, of two variables, represented by non-commuting operators $\hat{A}$ and $\hat{B}$, measured in the same state. The so-called Robertson's uncertainty relation \cite{Rob} reads
\begin{eqnarray} \label{q1}
\sigma_A\sigma_B\ge \frac{|\la[\hat{A},\hat{B}]\ra|}{2}, 
\end{eqnarray}
where $\sigma^2_{\hat{X}}\equiv \la\hat{X}^2\ra- \la\hat{X}\ra^2$, $[\hat{A},\hat{B}]\equiv \hat{A}\hat{B}-\hat{B}\hat{A}$ is the commutator, and the angular brackets indicate averaging in the pure state $|\Psi\ra$ the system is supposed to be in, $\la\hat{X}\ra=
\la\Psi|\hat{X}|\Psi\ra$. Equation (\ref{q1}) is often interpreted as an indication of the fact that a measurement of $\hat A$ must disturb a measurement of $\hat B$, and vice versa.
\newline
A somewhat different approach to the UP can be found in \cite{Feyn1}, where one reads:
{\it Any determination of the alternative
taken by a process capable of following more that one alternative
destroys the interference between alternatives}. 
%\newline
This principle is complemented by the rule for assigning probabilities to alternative scenarios \cite{Feyn2}:
{\it When an
event can occur in several alternative ways, the probability
amplitude for the event is the sum of the probability amplitudes
for each way considered separately. If an experiment is
performed which is capable of determining when one or another
alternative is actually taken, the probability of the event
is the sum of the probabilities of each alternative}.

From the above statements, to which we will refer as the "Feynman's uncertainty principle", one may conclude that interfering alternatives cannot be told
apart and must, therefore, form a single indivisible pathway \cite{DS3b}. Another corollary to the principle is that interference must be destroyed by a physical agent, e.g., a meter coupled to the observed system. Thus, construction of probabilities by means of squaring the moduli of the corresponding amplitudes, pre-supposes the existence of a suitable meter, as well as the fact that the meter has already been deployed \cite{flast}. 
\newline It is the Feynman's UP to which we will appeal in what follows. Despite the obvious difference between the formulations,  the Feynman's  principle is not all that different from the UP expressed by Eq.(\ref{q1}). Consider, for example, a system with a zero hamiltonian, $\hat{H}=0$, prepared at $t=0$ in a state $|\psi\ra$. An accurate  measurement of $\hat{B}$ at $t=T$ yields 
an eigenvalue $b_{i_B}$ with the probability $p(i_B|\psi)=|\la i_B|\psi\ra|^2$, where $|i_B\ra$ is the corresponding eigenstate. 
Suppose we also want to learn something about the value of $\hat{A}$ at $t=T/2$. The system must be, in some sense, in one of the eigenstates of $\hat{A}$. To see in what sense, precisely, we insert the unity $\sum_{i_A}|i_A\ra\la i_A|=1$ into the amplitude 
$\la i_B|\psi\ra$. The result is $\sum_{i_A}\la i_B|i_A\ra\la i_A|\psi\ra$. There are now different ways (paths, pathways, routes) to reach the final state by "passing 
through a state $|i_A\ra$. Without a meter, we cannot determine which of them is actually taken. Thus, despite the abundance of virtual paths, there is only a single "real" one, "travelled" with the probability $p(i_B)$. If, on the other hand, we employ a meter, capable of telling us which of the eigenvalues $a_{i_A}$ has occurred at $t=T/2$, the virtual paths will become real, with the $i_A$-th path travelled with the probability  $p(i_B|i_A|\psi)=|\la i_B|i_A\ra\la i_A|\psi\ra|^2$. We also find that the measurement of $\hat{A}$ has disturbed the result of measuring $\hat{B}$, since
\begin{eqnarray} \label{q2}
\sum_{i_A}p(i_B|i_A|\psi)\ne p(i_B|\psi). 
\end{eqnarray}
Quantifying the disturbance would lead to estimates similar to (\ref{q1}). We will not follow this matter any further, and in the next Section discuss possible realisation of quantum measurements. 
%%%%%%%%%%%%%%%%%%%%%%%%%%%%%%%%%%%%%%%%%%%%%%%%%%
\section {Von Neumann measurements and meters}
In the following we will often refer to von Neumann (vN) measurements \cite{vN}, whose definition we will revisit in this Section. A vN meter, designed to measure a variable $\hat{A}$ of a system governed by a hamiltonian $\hat{H}$,
consists of  a pointer (e.g., a massive particle on a line) whose position is $\xi$, and which is coupled to the observed system via ($\hbar=1$)
\begin{eqnarray} \label{n1}
\hat{H}_{int}=-i\beta(t)\frac{\partial}{\partial \xi} \hat{A}.
\end{eqnarray}
%so that the full Hamiltonian reads $\hat{H}+
If the value of $\hat{A}$ is measured at some $t=t_0$, the function $\beta(t)$ is chosen to be $\beta(t)=\delta(t-t_0)$, 
where $\delta(z)$ is the Dirac delta. We will assume that $\hat{A}$ has discrete eigenvalues $a_i$, and $|i\ra$ are the corresponding eigenstates, $\hat{A}=\sum_{i} |i\ra a_i\la i|$. Prior to the interaction, at $t=0$, the system is described by a state $|\psi\ra=\sum c_i(0)|i\ra$, and the pointer is in the state $|M\ra$, such that 
\begin{eqnarray} \label{n1a}
G(\xi)\equiv \la \xi|M\ra
\end{eqnarray}
 is a real function (e.g., a Gaussian),  peaked around $\xi=0$, and rapidly tending to zero as $\xi\to \pm \infty$  \cite{FOOTG}. Immediately after the interaction has taken place, the pointer becomes entangled with the system,  and their wave function is given by
\begin{eqnarray} \label{n2}
\la \xi|\Psi(t_0+0)\ra =\sum_{i} c_i(t_0) G(\xi -a_i)|i\ra,
\end{eqnarray}
where $c_i(t_0)=\la i |\exp[-i\hat{H}(t_0)]\psi\ra$
We assume further that the final position of the pointer can be determined precisely, leaving out the fundamental question of how exactly this is done, and whether a collapse of the wave function has occurred  \cite{vN}. In one way or another, after the interaction has taken place, we may "look" at the pointer, and find a meter's reading $\xi$. 
Perhaps the best known practical realisation of a vN measurement is the Stern-Gerlach apparatus (see, for example, \cite{Ah} and \cite{Bohm}). 
Other vN meters can be constructed by using the spin-orbit interaction \cite{SO1}- \cite{SO3}, or Bose-Einstein condensates trapped in double-well potentials \cite{DSprl}.
\newline
If nothing else is done to the system, 
the probability  (density) to find a reading $\xi_0$ is ($tr [...]$ stands for the trace of the operator inside the brackets)
\begin{eqnarray} \label{n3}
P(\xi_0)=tr [|\xi_0\ra\la\xi_0|\Psi(t)\ra\la\Psi(t)|]= \sum_i |c_i(t_0)|^2G^2(\xi_0-a_i),
\end{eqnarray} 
where $|\Psi(t)\ra =\exp[-i\hat{H}(t-t_0)]|\Psi(t+0)\ra$. Alternatively,  at some $t=T > t_0$ we may try to find (post-select) the system in some state $|\phi\ra$, and keep the meter's readings only if the post-selection is successful. With this additional condition the probability to find a reading $\xi_0$ becomes 
\begin{eqnarray} \label{n4}
P
%^{\phi \gets \psi}
(\xi_0)=tr [|\xi_0\ra\la\xi_0|\Psi(T)\ra\la\Psi|(T)|\phi\ra\la \phi|]=\n
 |\sum_i \la \phi |\exp[-i\hat{H}(T-t_0)]|i\ra G(\xi_0-a_i)\la i|\exp(-i\hat{H}t_0)|\psi\ra|^2,
\end{eqnarray}
\newline
Thus, apart from the preparation (pre-selection) of the system in a state $|\psi\ra$, two other events have occurred:
the pointer was found pointing at $\xi_0$, and the system was later found in a final state $|\phi\ra$. Quantum mechanics provides the probabilities of such occurrences by means of expressions similar to Eqs.(\ref{n3}) and (\ref{n4}). We can imagine, for example, performing more measurements between $t=0$ and $t=T$, thus creating a network of scenarios, each of which would occur with certain frequency if the experiment is repeated many times. Before discussing such "quantum" networks, we briefly turn to their classical counterparts. 
%%%%%%%%%%%%%%%%%%%%%%%%%%%%%%%%%%%%%%%%%%%%%%%%%%
\section {Classical networks and functionals}
We start with a simple toy model which, we hope, will be helpful for the discussion of the following Sections.
Suppose one has at his/her disposal a kit containing purely classical elements, little balls,  tubes and connectors,
which can be joined together. A connector $x$ has two entrances,  and $w_x(i|j)$ is the probability to enter via the inlet $j$, and subsequently  leave via the outlet $i$. If one of the outlets is blocked, the probability to exit via the remaining one is unity for both entrances. Connecting the elements as 
shown in Fig. 1, one can  "engineer" a network of paths, which a ball inserted at the top will travel at random, until ending up in one of the final states $f_1$ and $f_2$. Ascribing variables (numbers)  $a_1$, $a_2$, $b_1$, $b_2$ to each of the connectors, one can also associate numbers, (functionals) with each path. For example, in Fig.1a there are eight paths $\{i\}$, $i=1,...,8$. Four of them end in the final state $f_1$,

$\{1\}$: $f_1 \gets b_1 \gets a_1 \gets in$, travelled with the probability $p[1]=w_{b_1}(1|1)w_{a_1}(1|1)w_{in}(1|1)$,

$\{2\}$: $f_1 \gets b_1 \gets a_2 \gets in$, travelled with the probability $p[2]=w_{b_1}(1|2)w_{a_2}(2|1)w_{in}(2|1)$,

$\{3\}$: $f_1 \gets b_2 \gets a_1 \gets in$, travelled with the probability $p[3]=w_{b_2}(2|2)w_{a_1}(2|1)w_{in}(1|1)$,

$\{4\}$: $f_1 \gets b_2 \gets a_2 \gets in$, travelled with the probability $p[4]=w_{b_2}(2|1)w_{a_2}(1|1)w_{in}(2|1)$,
and another four,  $\{j\}$, $j=5,6,7,8$, end in $f_2$. It is always possible to consider sub-networks containing only few paths, regardless of what is happening at later times, or elsewhere in the diagram. 
\newline
For example, we may limit our attention only to the cases where the ball goes to $b_1$. As a result of this "post-selection" there are only two paths: $\{1\}$, through $b_1\gets a_1 \gets  in$ and 
$\{2\}$, through $b_1\gets a_2 \gets  in$, travelled with probabilities $p[1]=w_{a_1}(1|1)w_{in}(1|1)$  and $p[2]=w_{a_1}(2|2)w_{in}(2|1)$, respectively. There is a single random variable, a functional  $F_1[i]$ taking the values $a_i$, $i=1,2$, and whose mean value, evaluated over many runs by counting the times the ball passes through $a_1$ and $a_2$, is
%\begin{eqnarray} \label{m1}
$\la F_1 \ra = \{ p[1]F[1]+p[2]F[2]\}/\{ p[1]+p[2]\}$.
%\end{eqnarray}
\newline Alternatively, we may focus only on those cases where the ball ends up in $f_1$, and ask for the  difference
between the values of $b_i$ and $a_j$, $b_i-a_j$, which is recorded for each run.  Let $a_1=b_1=-1$ and $a_2=b_2=1$. Now the 
said difference is a functional $F_2[path]$ defined on the four paths introduced above, and taking the values
$F_2[1]=F_2[4]=0$, $F_2[2]=-2$ and $F_2[3]=2$. Since two of the values of $F_2$ are the same, we can reduce the number of paths by one, by combining $\{1\}$ and $\{4\}$ into a single pathway, $\{1+4\}$, travelled with a probability $p[1+4]=p[1]+p[4]$, 
and on which $F_2[1+4]$ is understood to be zero. The resulting equivalent system is shown in Fig.1b, and for $\la F_2\ra$,
evaluated by writing down the value of $F_2$ for each run of the ball, adding up the results, and dividing by the number of runs, we find
\begin{eqnarray} \label{m1}
\la F_2 \ra = \frac{p[2]F[1]+p[3]F[2]}{p[1+4]+p[2]+p[3]}.
\end{eqnarray}
\begin{figure}[ht]
\epsfig{file=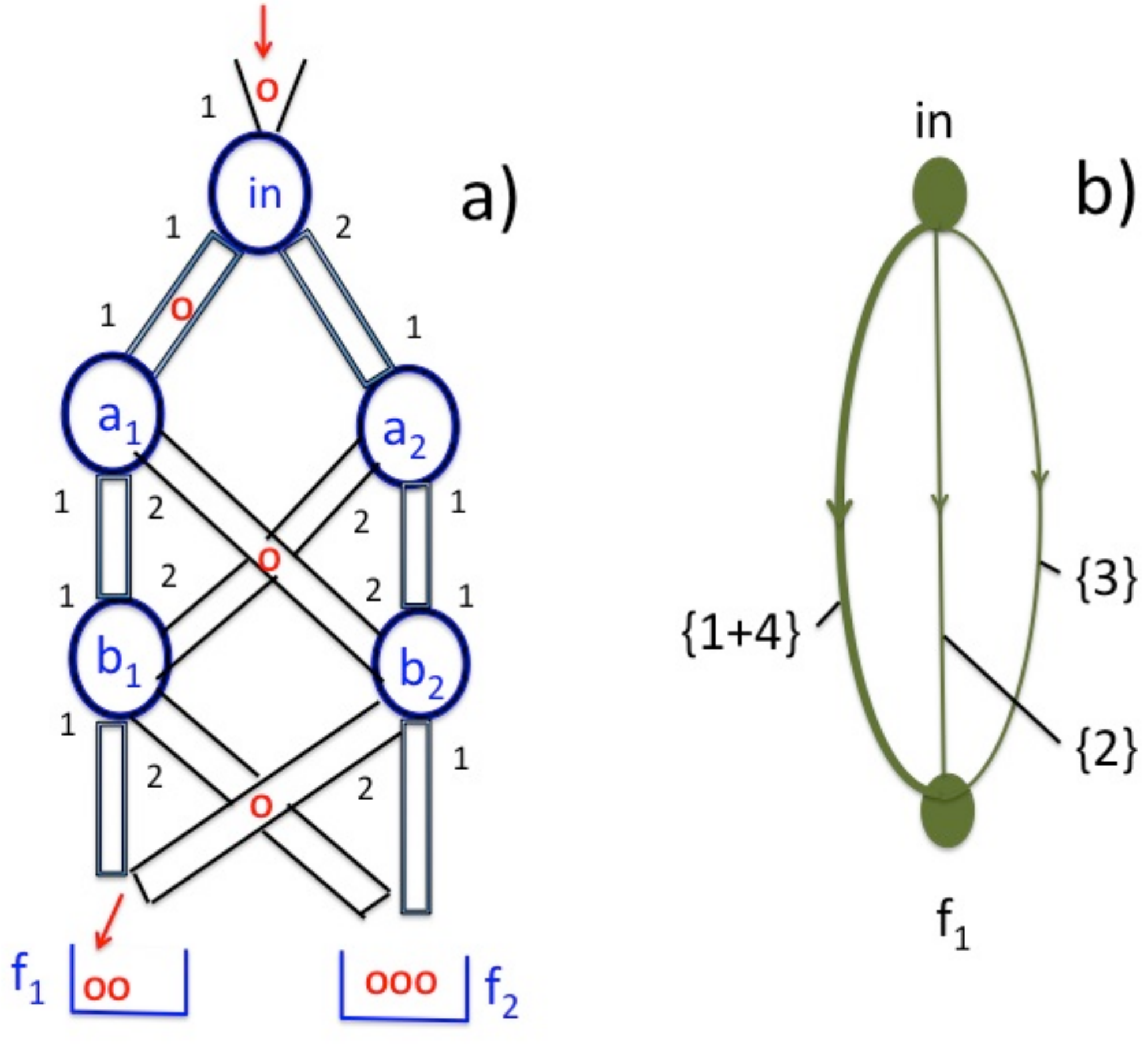, height=8cm, width=12cm, angle=0}
\vspace{0pt}
\caption{
a) Schematic diagram of the toy model of Sect. IV. A ball is introduced at the top and follows the path $\{3\}$ before ending up in the receptacle $f_1$. The numbers mark the inlets and outlets of a connector, with the closed ones not shown. b) A reduced model 
for the system post-selected in $f_2$, with only $F_2$ measured.}
\end{figure}
We have invoked the simple toy model in Fig.1 in order to emphasise its resemblances to and differences from the quantum case outlined at the end of previous Section. As in the classical case, a set of vN measurements creates sequences of events occurring with 
certain probabilities, which can be seen as classical pathways "travelled" with certain frequencies, if the the experiment is repeated many times. 
\newline
The quantum origin of the experiment becomes evident in the way the probabilities of the events are calculated, 
as well as in the nature of these events. A meter recombines and divides the system's  virtual paths, endowed with probability amplitudes only, thereby producing "real" ones, to which probabilities can be assigned. In particular, two interfering paths are combined into a single one 
in a much stronger sense, than in the classical case. For example, in Fig.1b  the paths $\{1\}$ and $\{4\}$ are combined into $\{1+4\}$, since we chose to be interested only in the difference $b_i$ of $a_i$, and not in their actual values.
We will show that in the quantum case, these actual values may not even exist. Finally, in the classical case, it is understood that two different setups 
built from the elements of the same kit can be completely different, and need not share any statistical properties. 
In the quantum case this is less clear, and we will return to the problem of "contextuality" later.
%%%%%%%%%%%%%%%%%%%%%%%%%%%%%%%%%%%%%%%%%%%%%%%%%%
\section {A simple quantum network: measuring a quantity with post-selection}
As the first example, we consider a vN measurement of an operator $\hat{A}=\sum_{i=1}^2 |i\ra a_i \la i|$, for a two-level system, pre- and post-selected in the states $|\psi\ra$ and $|\phi\ra$, at $t=0$ and $t=T$, respectively. For simplicity we put the system's hamiltonian to zero, $\hat{H}=0$, so that nothing happens between $t=0$ and $t=t_0$, as well as between $t=t_0$ and $t=T$.
Our purpose is to show that this creates a classical statistical ensemble not too dissimilar from the one built from the classical elements in the previous Section. To see how this is done, we consider the two {\it virtual} paths connecting $|\psi\ra$ and $|\phi\ra$, $\{i\}$, $i=1,2$.
These are  

$\{1\}$: $|\phi\ra \gets  |1\ra \gets |\psi\ra$, with an amplitude $A[1]=\la \phi |1\ra\la1|\psi\ra$, \q and

$\{2\}$: $|\phi\ra \gets |2\ra \gets |\psi\ra$, with an amplitude $A[2]=\la \phi |2\ra\la 2|\psi\ra$.
\newline
We also have $\sum_{i=1}^2A[i]=\la \phi|\psi\ra$.

The value of $\hat{A}$ at $t=t_0<T$ is a functional $F_1[path]$ defined on the two paths and, obviously, taking the values $a_1$ and $a_2$
\begin{eqnarray} \label{y2}
 F_1[1]=a_1,\q \text{and} \q F[2]=a_2.
\end{eqnarray}
Without a meter, we may define an {\it amplitude distribution} $\Phi(f)$ for the value $f$ taken by $F_1$, 
\begin{eqnarray} \label{y3}
\Phi(f)=A[1]\delta(f-a_1)+A[2]\delta(f-a_2), \q \int \Phi(f)df=\sum_{i=1}^2A[i]=\la \phi|\psi\ra 
\end{eqnarray}
but cannot say what is the {\it probability} for having, e.g., the value $a_1$. 
\newline
With a meter (\ref{q1}) coupled to the system, the final pointer's state $|M'\ra$ is easily found to be given by
\begin{eqnarray} \label{y4}
M'(\xi) \equiv \la \xi|M'\ra
=\int G(\xi-f)\Phi(f)df=
%\q\q\q\q\q\q\n
A[1]G(\xi-a_1)+A[2]G(\xi-a_2),\q\q
\end{eqnarray}
and the meter reads $\xi_0$ with a (unnormalised) probability 
\begin{eqnarray} \label{y5}
P(\xi_0)=|M'(\xi)|^2=|A[1]G(\xi_0-a_1)+A[2]G(\xi_0-a_2)|^2. 
\end{eqnarray}
Equation (\ref{y5}) may be interpreted in the following way. By turning the meter on, we have created a continuum of {\it real} paths, all labelled by $\xi$. These connect the three observed events, $system\q in\q|\phi\ra$ $\gets$ $pointer\q at\q \xi_0$ $\gets$ $system\q in\q|\psi\ra$, 
assuming the meter is read before the post-selection is performed.
\newline
The range of possible scenarios is determined by the width of the initial pointer's state $G(\xi)$ in Eq.(\ref{n1a}), which also determines the accuracy of the measurement. For example, if $G(\xi)$ is chosen to be 
\begin{eqnarray} \label{y5b}
G(\xi)=1/\sqrt{\Delta \xi}\q \text {for} \q |\xi|\le \Delta \xi/2, \q \text{and}\q 0 \q \text{otherwise}, 
\end{eqnarray}
% and $0$ otherwise, 
 the possible values of $\xi$ would lie inside the the interval $[a_1-\Delta \xi/2, a_2+\Delta \xi/2]$, assuming $a_2>a_1$.
The path labelled $\xi$ is travelled with a probability (density) $P(\xi)$. Far from passively observing the motion of the system, a meter actively participates in shaping the observed statistical ensemble.
Accordingly, the probabilities in Eq.(\ref{y5}) contain both the amplitudes $A[i]$, which characterise the unobserved system, and the meter's own $G(\xi)$.
\newline
In the general case we may learn little about the the quantity of interest, $\hat{A}$. Instead of extracting its value, we can only conclude that if 
the meter reading is $\xi_0$, 
%after the coupling (\ref{y1}),
%has been applied, 
the system's state after interaction with the meter was \cite{FOOT14}
\begin{eqnarray} \label{y6}
%|\Psi(t_1+0)\ra=G(\xi_0-1)\la+|\psi\ra|+\ra+G(\xi_0+1)\la-|\psi\ra|-\ra 
|\Psi(\xi_0)\ra=G(\xi_0-a_{1})\la1|\psi\ra|1\ra+G(\xi_0-a_{2})\la 2|\psi\ra(\xi_0)|2\ra,
% \q \alpha_{1,2}=G(\xi_0-a_{1,2})\la1,2|\psi\ra,
\end{eqnarray}
for which the precise value of $\hat{A}$ remains indeterminate. It is, however, possible to say that the measurement "drives" the system 
through a state $|\Psi(\xi_0)\ra$, with a probability $P(\xi_0)$. To see that this is, indeed, the case, we could make an accurate measurement of the projector onto $|\Psi(\xi)\ra$ just after $t_0$. Repeating the procedure many times will show that whenever the reading 
of the first meter lies in a narrow interval around $\xi_0$, the additional projection will also succeed. 
Thus, while every attempt to measure $\hat{A}$  leads to a particular stochastic network, not every measurement allows us to determine the value of $\hat{A}$. We will return to the  question of the accuracy shortly after considering yet another example. 
\begin{figure}[ht]
\epsfig{file=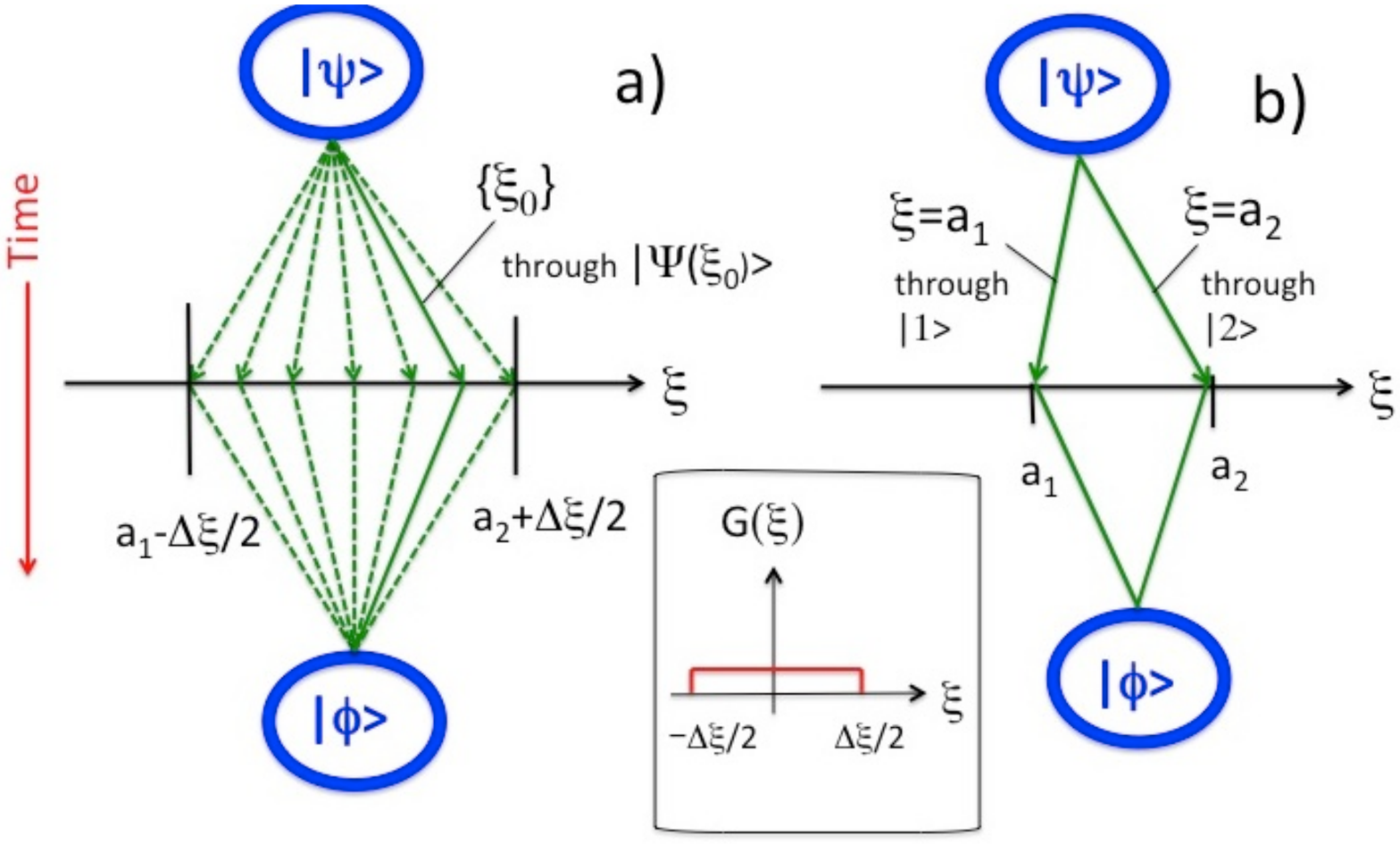, height=9cm, width=18cm, angle=0}
\vspace{0pt}
\caption{
a) Schematic diagram showing a continuum of real pathways produced in a measurement of an operator  $\hat{A}$ with eigenvalues $a_1<a_2$.  A pathway connects three observable events: preparation in  $|\psi\ra$ at $t=0$, obtaining  a meter reading 
$\xi_0$ at $t=T/2$, and finding the system in $|\phi\ra$ at $t=T$. If a reading $\xi_0$ obtained, the system "passes" through the state $|\Psi(\xi_0)\ra$ in Eq.(\ref{y6}). For $G(\xi)$ in Eq.(\ref{y5b}) (see the inset), the pointer readings are contained in the interval 
$[a_1-\Delta \xi/2,a_2+\Delta \xi/2]$.  b) The two real pathways surviving in the high accuracy limit, $\Delta \xi \to 0$.}
\end{figure}
%%%%%%%%%%%%%%%%%%%%%%%%%%%%%%%%%%%%%%%%%%%%%%%
\section {A more elaborate network: measuring the difference of two quantities}
%, $\hat{B}(t_2)-\hat{A}(t_1)$}
Suppose that, as in previous Section, we have a two-level system pre- and post-selected in the states 
$|\psi\ra$ and $|\phi\ra$ at $t=0$ and $t=T$, respectively. Now we want to learn something about the difference
of the quantities represented by operators $\hat{A}=\sum_{i_A=1}^2 |i_A\ra a_{i_A} \la i_A|$ and 
$\hat{B}=\sum_{i_B=1}^2 |i_B\ra b_{i_B} \la i_B|$, which may or may not commute,  at $t=t_1$ and $t=t_2$, such that $0< t_1<t_2<T$.
[Classically, we could consider, for example, the $x$-and $y$-components of the momentum, and ask for $p_y(t_2) -p_x(t_1)$]. Quantally, there is more than one way to approach this task.

(i) We could employ two vN meters, set to measure $\hat{A}$ and $\hat{B}$ separately, and evaluate the difference of their readings.

(ii) We could also construct a hermitian operator formally representing the difference of interest, $\hat{C}=\exp[-i\hat{H}(t_2-t_1)]
\hat{B}\exp[i\hat{H}(t_2-t_1)]-\hat{A}$, and use a single vN meter to measure it at $t=t_1$. As discussed in \cite{DSproc},
a classical analog of this procedure is a predictive measurement, which uses the fact that knowing the system's coordinates and momenta at $t=t_1$ allows one to restore the system's subsequent trajectory and, with it, all future values of all quantities.

(iii) A procedure unique to quantum mechanics allows one to evaluate the difference of interest, leaving the actual values of $\hat{A}$ and $\hat{B}$ indeterminate. 
\newline
In general, all three methods give different results \cite{DSproc}, and here we are only interested in the option (iii). 
With this in mind, we couple a vN pointer (position $\xi$), to $\hat{A}$ and $\hat{B}$ at $t=t_1$ and $t=t_2$, respectively, so that the interaction with the meter is given by 
\begin{eqnarray} \label{x1}
H_{int}= -i\frac{\partial}{\partial \xi}[\hat{B}\delta(t-t_2)-\hat{A}\delta(t-t_1)],
\end{eqnarray}
and the pointer's initial state, $G(\xi)$ has, as before, a Gaussian shape, centred at $\xi=0$. The measurement works as follows: after the first coupling the pointer is shifted by one of the eigenvalues of $\hat{A}$, and after the second one by an eigenvalue of $\hat{B}$, albeit in the opposite sense. The resulting shift is, thus, the difference of the two values. This is not exactly the scheme used by  von Neumann \cite{vN}, and we will call this meter von Neuman-like \cite{f3}.
%Let $\hat{A}=\sigma_x$ and $\hat{B}=\sigma_y$ be two non-commuting components of the system's spin. To simplify the formulae, we may also put $\hat{H}=0$, so that nothing happens
%unless the system is coupled to the meter. Both $\hat{A}$ and $\hat{B}$ have eigenvalues $\pm 1$, and we will denote the corresponding eigenstates $|\pm_x\ra$ and $|\pm_y\ra$.

Without a meter there are four virtual paths, $\{i\}$, $i=1,2,3,4$, which connect $|\psi\ra$ and $|\phi\ra$, and pass through different intermediate states. For example, the path going through $|1_A\ra$, and then through $|2_B\ra$ has a probability amplitude
$\la \phi |\exp[-i\hat{H}(T-t_2)] |2_B\ra\la2_B|\exp[-i\hat{H}(t_2-t_1)]|1_A\ra\la1_A|\exp(-i\hat{H}t_1)]|\psi\ra$.
Putting, as before, $\hat{H}=0$ we, therefore,  have the paths

$\{1\}$: $|\phi\ra \gets |1_B\ra \gets |1_A\ra \gets |\psi\ra$, with an amplitude $A[1]=\la \phi |1_B\ra\la1_B| 1_A\ra\la1_A|\psi\ra$, 

$\{2\}$: $|\phi\ra \gets |1_B\ra \gets |2_A\ra \gets |\psi\ra$, with an amplitude $A[2]=\la \phi|1_B\ra\la1_B| 2_A\ra\la2_A|\psi\ra$,

$\{3\}$: $|\phi\ra \gets |2_B\ra \gets |1_A\ra \gets |\psi\ra$, with an amplitude $A[3]=\la \phi|2_B\ra\la2_B| 1_A\ra\la1_A|\psi\ra$,

$\{4\}$: $|\phi\ra \gets |2_B\ra \gets |2_A\ra \gets |\psi\ra$, with an amplitude $A[4]=\la \phi|2_B\ra\la2_B| 2_A\ra\la2_A|\psi\ra$,
\newline
and also the condition $\sum_{i=1}^4A[i]=\la \phi|\psi\ra$. The four paths are easily identified in the diagram in Fig.3a.
\newline
The difference between $\hat{A}$ and $\hat{B}$ is a functional $F_2[path]$ defined on the four paths, where it may take four different values
\begin{eqnarray} \label{x2}
F_2[1]=b_1-a_1,\q F_2[2]=b_1-a_2,\q F_2[3]=b_2-a_1,\q \text{and} \q F_2[4]=b_2-a_2.
\end{eqnarray}
Let $\hat{A}$ and $\hat{B}$ represent components of the spin (without a factor of 1/2)  along two different axes, so that $a_1=b_1=-1$, and  $a_2=b_2=1$. Now $F_2$ takes three different values, $-2$, $0$ and $2$.
With no meter employed, we cannot say what is the chance of having, e.g., $F=2$, but, as before, may write down an amplitude distribution $\Phi(f)$ for the value $f$ of the functional $F_2$, 
\begin{eqnarray} \label{x3}
\Phi(f)=\{A[1]+A[4]\}\delta(f)+A[2]\delta(f+2)+A[3]\delta(f-2),
\end{eqnarray}
normalised by  the condition $\int \Phi(f)df=\sum_{i=1}^4A[i]=\la \phi|\psi\ra$.

As in the previous Section, this so far purely cosmetic re-arrangement of the virtual paths is useful for describing the effects of a meter, once it is coupled to the system
as prescribed by Eq. (\ref{x1}). After a successful post-selection the pointer is in a pure state $|M'\ra$, and it is a simple matter to check 
that 
\begin{eqnarray} \label{x4}
M'(\xi) \equiv \la \xi|M'\ra
=\int G(\xi-f)\Phi(f)df=
%\q\q\q\q\q\q\n
 \{A[1]+A[4]\}G(\xi)+A[2]G(\xi+2)+A[3]G(\xi-2).\q\q
\end{eqnarray}
We can now look at Eq.(\ref{x4}) from the point of view adopted in the previous Section.
As in Sect. IV, by external manipulation (connecting the system to a meter) we have created a classical statistical ensemble, 
where the system can reach the final state vie a continuum of real paths, labelled by the the variable $\xi$.
Again, the meter plays an active role, and the (unnormalised) probability to travel the $\xi_0$-th path is determined by the shape of $G$,
which is the property of the pointer, as well as the amplitudes of the virtual paths, which represent the system "on its own",
\begin{eqnarray} \label{x5}
%\la F_2 \ra = \frac{p[2]F[1]+p[3]F[2]}{p[1+4]+p[2]+p[3]}.
P(\xi_0)=| \{A[1]+A[4]\}G(\xi)+A[2]G(\xi+2)+A[3]G(\xi-2)|^2.
\end{eqnarray}
With this, the mean reading of the meter over many trials is readily seen to be given by 
\begin{eqnarray} \label{x6}
\overline{\xi}=\frac{\int \xi P(\xi)d\xi}{\int P(\xi)d\xi}.
\end{eqnarray}
As in the previous example, the value of $F_2$ remains indeterminate, if there is more than one term in r.h.s. of Eq.(\ref{x5}).
To establish a good correlation between the pointer readings $\xi$ and the values of $F_2$ we must, therefore, try to avoid the overlap between 
different $G$'s in (\ref{x5}). We will do so after discussing the general rules for combining virtual paths in the next Section.
%%%%%%%%%%%%%%%%%%%%%%%%%%%%%%%%%%%%%%%%%%%%%%%%%%%%
\begin{figure}[ht]
\epsfig{file=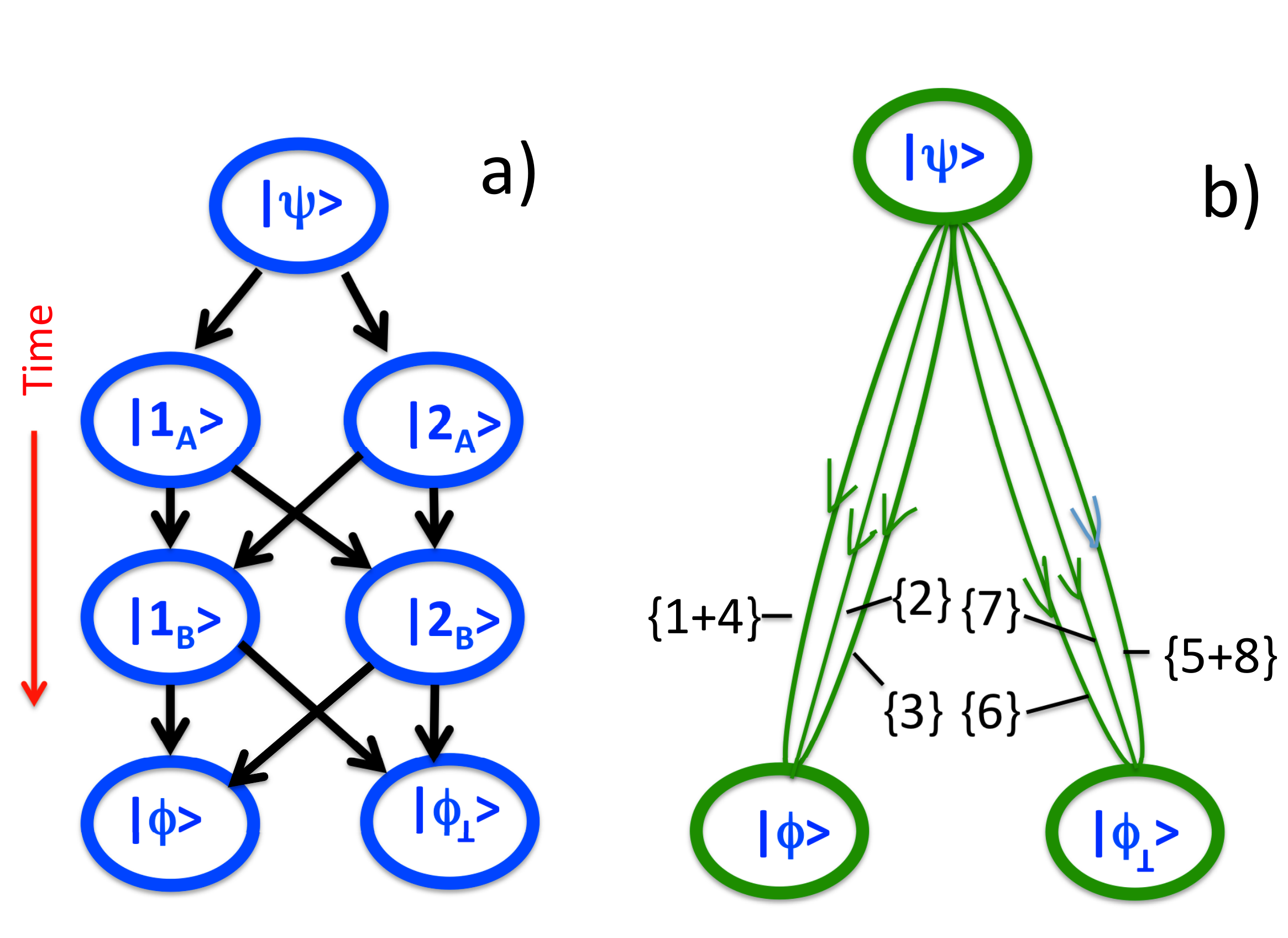, height=7cm, angle=0}
\vspace{0pt}
\caption{
a) Eight virtual paths connecting the  initial state of a two-level system, $|\psi\ra$, with its final state $|\phi\ra$, and its orthogonal compliment $|\phi_\perp\ra$. b) Real paths produced in an accurate measurement of the functional $F_2$ in Eq.(\ref{x2}), which takes the values $-2$, $0$, and $2$.}
\end{figure}
%%%%%%%%%%%%%%%%%%%%%%%%%%%%%%%%%%%%%%%%%%%%%%%%%%%%%%%%
\section {Superposition principle for virtual paths}
The Feynman's uncertainty principle, cited in Sect. II, implies that virtual paths can be combined and recombined into new sets of paths, with interference being responsible for the loss of information about certain aspects of quantum motion.
Next we briefly discuss the rules for creating such combinations.
\newline
We start with the well known case of changing the basis functions in which a state of the system can be expanded at a given time.
Let a two-state system be in the state $|\Psi\ra=a_1|\varphi_1\ra+a_2|\varphi_2\ra$, where $\la \varphi_m|\varphi_{m'}\ra=\delta_{mm'}$, and $a_m$ is the amplitude with which $|\varphi_m\ra$ enters in $|\Psi\ra$. Consider a different basis, $|\phi_i\ra$, $\la \phi_i|\phi_{i'}\ra=\delta_{ii'}$, which is related to the first basis by a unitary transformation, $|\varphi_i\ra=u_{i1}|\phi_1\ra+u_{i2}|\phi_2\ra$. Now the amplitude $b_i$, with which $|\phi_i\ra$ enters in $|\Psi\ra$, is
%\begin{eqnarray} \label{z1}
$b_i=\la \phi_i|\Psi\ra=\sum_j u_{ji}a_j$, $i=1,2$.
%\end{eqnarray}
These equations can be read as follows: if a new basis state is a weighted sum of the old ones, the corresponding amplitude is the weighted sum of the corresponding amplitudes.
\newline
The same rules can be applied to virtual paths, and we can describe the action of the meter in Sect.V in a similar manner.
If the pointer reading is $\xi_0$, the meter combines virtual paths $\{1\}$ and $\{2\}$ with amplitudes $A[1]$ and $A[2]$, into a single real path
$G(\xi_0-a_1)\{1\}+G(\xi_0-a_2)\{2\}$, *h the amplitudes $A[1]$ and $A[2]$, are combined into a new path, $\alpha \{1\}+\beta \{2\}$, with complex valued weights $\alpha$ and $\beta$, then the amplitude for the new pathway is
\begin{eqnarray} \label{z3}
A[\alpha \{1\}+\beta \{2\}]=\alpha A[1] + \beta A[2].
\end{eqnarray}
If a functional $F[i]$ takes different definite values on $\{1\}$ and $\{2\}$, $F[1]\ne F[2]$, its value on $\alpha \{1\}+\beta \{2\}$ is indeterminate, 
 (we will elaborate on this statement in Sect.X).
\newline
That  the width of $G(\xi)$, $\Delta \xi$, determines the resolution, or the accuracy,  of a measurement is readily seen if $G(\xi)$ is chosen
to have the form of a "rectangular window' (\ref{y5b}). Inserting (\ref{y5b}) in Eq. (\ref{x4}) yields 
\begin{eqnarray} \label{z4}
P(\xi_0)=\Delta \xi^{-1}|\int_{\xi_0-\Delta \xi/2}^{\xi_0+\Delta \xi/2}\Phi(f)df|^2,
\end{eqnarray}
and we can be sure that if the reading if $\xi_0$, the value of the measured functional, $f$, is no smaller than $\xi_0-\Delta \xi/2$, and no greater than $\xi_0+\Delta \xi/2$. We are, however, "banned" from looking inside the interval $[\xi_0-\Delta \xi/2, \xi_0+\Delta \xi/2]$, 
where all information about $f$ is lost to the interference which the meter has failed to destroy. 
Next we look at the question of accuracy in more detail.
%%%%%%%%%%%%%%%%%%%%%%%%%%%%%%%%%%%%%%%%%%%%
\section{Most accurate measurements, real pathways and contextuality.}
Suppose we can change the width $\Delta \xi$ of the initial pointer state, e.g., by choosing 
\begin{eqnarray} \label{w1}
\la \xi|M\ra\equiv G(\xi|\Delta \xi)= (\Delta \xi)^{-1/2}G(\xi/\Delta \xi),
\end{eqnarray}
where the factor $(\Delta \xi)^{-1/2}$ ensures that the state is properly normalised, $\int |G(\xi|\Delta \xi)|^2d\xi=1$.
Since, by assumption, the  pointer's final position is evaluated accurately, it is the uncertainty of its initial setting 
that determines precision of the measurement. For a highly accurate measurements we should choose $\Delta \xi$ as small as possible, thus requiring 
\begin{eqnarray} \label{w1a}
\Delta \xi\to 0, \q |G(\xi|\Delta \xi)|^2\to \delta(\xi).
\end{eqnarray}
If so, in the example of Sect. V, Eq.(\ref{y5}) reduces to
\begin{eqnarray} \label{w25}
P(\xi)=|A[1]|^2 \delta(\xi-a_1)+|A[2]|^2 \delta(\xi-a_2)\equiv p[1] \delta(\xi-a_1)+p[2] \delta(\xi-a_2). 
\end{eqnarray} 
and of all real paths survive only two. Equivalently, we can say that 
the action of the meter has destroyed all interference between the virtual paths $\{1\}$ and $\{2\}$, 
which can now be equipped with the probabilities $p[1]$ and $p[2]$, respectively.
The paths are "real" in the sense that each time the experiment is repeated, the pointer will indicate which of 
the two paths has been travelled, with $p[1]/(p[1]+p[2])$ and $p[2]/(p[1]+p[2]$ giving the relative frequencies of each occurrence. 
Note that the appearance of $|A[i]|^2$, later interpreted as probabilities, is not postulated,
 but is a consequence of the change in the physical state of the system, caused by its interaction with the meter \cite{flast}.
\newline
Recalling that whenever post-selection in the state $|\phi\ra$ fails, the system ends up in its orthogonal compliment, 
$|\phi_\perp\ra$, $|\phi\ra\la\phi|+|\phi_\perp\ra\la\phi_\perp|=1$, we can add two new paths,  $\{3\}$ and $\{4\}$,  connecting $|\psi\ra$ with 
$|\phi_\perp\ra$, and travelled with probabilities
\begin{eqnarray} \label{w3}
p[3]=|\la \phi_\perp |1\ra\la1|\psi\ra|^2, 
\end{eqnarray}
and 
\begin{eqnarray} \label{w4a}
p[4]=|\la \phi_\perp |2\ra\la2|\psi\ra|^2. 
\end{eqnarray}
respectively. We now have an exact equivalent of the (upper part of the) model in Fig.1. The only difference is that this time it is built from quantum "elements", rather than from pieces of a classical kit. The rest is a simple exercise in classical statistics. 
 In both cases we can calculate the averages of the functional $F_1$.
For example, conditioned by successful post-selection in $|\phi\ra$, the mean value of $F_1$ is given by
$\la F_1\ra =(F_1[1]p[1]+F_1[2]p[2])/(p[1]+p[2])$. We can also choose the operator $\hat{A}$ to be the projector on the state $|1\ra$, so that its eigenvalues are $1$ and $0$, and we have $F_1[1]=1$ and $F_1[2]=0$. Now the mean value of $F_1$ yields precisely the probability to travel the first path, 
\begin{eqnarray} \label{w4}
\la F_1\ra=\frac{p[1]}{p[1]+p[2]}.
\end{eqnarray}
Classically, we would arrive at the same result by writing down $1$ whenever the ball passes through $a_1$, $0$ if it passes through $a_2$, add up what is written, and divide by the number of the balls we used.
\newline
The case case of measuring the difference of operators $\hat{B}(t_2)$ and $\hat{A}(t_1)$, discussed in Sect.VI, can be analysed in a similar manner.
There an accurate meter creates three real paths connecting $|\psi\ra$ and $|\phi\ra$. The three pathways are 

$\{1+4\}$, travelled with a probability $p[1+4]=|A[1]+A[4]|^2$, 

$\{2\}$, travelled with a probability $p[2]=|A[2]|^2$, and

$\{3\}$, travelled with a probability $p[3]=|A[3]|^2$.
\newline
Adding three more paths connecting $|\psi\ra$ with $|\phi_\perp\ra$, we obtain a classical stochastic system shown in Fig.3,
similar to that built from the tubes and connectors in Fig.1.
The mean value of the functional $F_2$ defined in Eq.(\ref{x2}), conditioned by arriving in $|\phi\ra$, is now twice the difference of the relative frequncies with which the paths $\{2\}$ and $\{3\}$ are travelled,
\begin{eqnarray} \label{w5}
\la F_2\ra=\frac{F_2[1]p[1+4]+F_2[2]p[2]+F_2[3]p[3]}{p[1+4]+p[2]+p[3]}= \frac{2(p[2]-p[3])}{p[1+4]+p[2]+p[3]}.
\end{eqnarray}
%We conclude with a remark following from what has just been said. 
To conclude, we return to the remark made at the end of Sect. IV.
Joining the connectors and tubes in different ways, one may construct completely different statistical networks, which have nothing 
in common, except their constituent parts. 
In a similar way, by applying different measurement schemes, and measuring different operators to different accuracies, one may engineer equally different stochastic ensembles, having nothing in common except their parent quantum system. Ignoring this fact
may lead to various counterfactual paradoxes, some of which disappear once the above remark is taken into account (for more detail see  \cite{PLA16} and the Refs. therein).
%%%%%%%%%%%%%%%%%%%%%%%%%%%%%%%%%%%%%%%%%%%%%%%%%%
\section{Most inaccurate  measurements and the meaning of "indeterminate"}
Next we look at what happens in the opposite limit, 
\begin{eqnarray} \label{v1}
\Delta \xi \to \infty.
\end{eqnarray}
There are several reasons for doing it. Firstly, as in the case of the accurate limit considered in the previous Section, 
the result is universal in the sense that it no longer depends on the particular choice of $G(\xi)$, and appears 
to represent some "intrinsic" property of the studied system. Secondly, observation of the system's history changes,
%, for example, 
%the probability to end up in the state $|\psi\ra$ since,
 in general, the probability to end up in the state $|\phi\ra$ which, when an accurate meter is employed, is different from that without a meter,
\begin{eqnarray} \label{v1a}
\sum_i p[i]\ne|\la \phi |\exp(-i\hat{H}T)|\psi\ra|^2.
\end{eqnarray}
This is a well known example of "one measurement  perturbing the other", and it is tempting to try to get rid of the perturbation, and to see what will then occur.
\newline
The perturbation disappears if, and and only if, all $G$'s in Eq.(\ref{x4}) have approximately the same value, 
$G(\xi) \approx G(\xi+2)\approx G(\xi-2)$, so that $P(\xi)$ in Eq.(\ref{x5}) is $\approx |G(\xi)|^2\la \phi |\exp(-i\hat{H}T)|\psi\ra|^2$.
This can only happen in the limit (\ref{v1}), when $G(\xi)$ is made very broad.
One immediate consequence of taking the limit (\ref{v1}) is that  the range of possible meter's readings has widened to 
occupy the whole real axis, 
\begin{eqnarray} \label{v2}
-\infty\le \xi \le \infty.
\end{eqnarray}
This gives an operational meaning to the word "indeterminate" we have often used above. If one wishes to keep interference between paths intact, he/she {\it must} choose the meter so inaccurate, that its reading will be arbitrary, leaving the researcher completely ignorant of the value of the functional in question.
\newline
Since it appears that not much can be learnt from an individual inaccurate measurement and, we follow \cite{Ah} and move on to look at the {\it average} meter reading, $\overline{\xi}$, obtained  as the number of trials $N$ tends to infinity. Nowt $\overline{\xi}$ must be expressed in the terms of the path amplitudes $A[i]$. The authors of  \cite{Ah} have shown that in the case of Sect.V the correct expression is
\begin{eqnarray} \label{v3}
\overline{\xi} =\text{Re}\frac{\sum_i a_i\la \phi|i\ra\la i| \psi\ra}{\sum_i \la \phi|i\ra\la i| \psi\ra}
=\text{Re}\frac{\sum_i F_1[i]A[i]}{\sum_i A[i]}=
\text{Re}\{a_1\alpha[1]+a_2\alpha[2]\}.
\end{eqnarray}
where 
\begin{eqnarray} \label{v4}
\al[i]=A[i]/\sum_i A[i]
\end{eqnarray}
is the {\it relative amplitude} for the path number $i$.
In the special case of $\hat{A}$ being the projector onto the state $|1\ra$, $\hat{A}=|1\ra\la1|$, we have
\begin{eqnarray} \label{v3a}
\overline{\xi} = \text{Re}\al[1],
\end{eqnarray}
In \cite{f3} the result was extended to any real functional for which, in the absence of a meter, 
the amplitude for taking a value $f$ is $\Phi(f)$,  
\begin{eqnarray} \label{v5}
\overline{\xi} =\text{Re}\frac{\int f \Phi(f) df }{\int \Phi(f) df}.
\end{eqnarray}
%Expression for the variance, $\sqrt{\xi^2-\overline{\xi}^2 }$, which can be found in \cite{NEGAT}, takes 
Similarly, the average reading of a very inaccurate meter measuring the difference  of $\hat{B}(t_2)$ and $\hat{A}(t_1)$ in the example of Sect.VI,
is given by 
\begin{eqnarray} \label{v6}
\overline{\xi} =\text{Re}\frac{F[1]A[1+4]+F[2]A[2]+F[3]A[3]}{A[1+4]+A[2]+A[3]}= 2\text{Re}(\al[2]-\al[3])
\end{eqnarray}
In general,  an average reading of a very inaccurate meter is expressed in terms of the relative amplitudes for the virtual paths connecting the initial and final states.
Equations (\ref{v3a}) and (\ref{v6}) bear uncanny resemblance to Eqs.(\ref{w4}) and (\ref{w5}), in which the probabilities $p[i]$ have been replaced by the corresponding probability amplitudes $A[i]$, and the real part was chosen to make the result real valued. This may be taken as a hint that the quantity under the $Re$ sign, the {\it weak value} of a functional $F$,
\begin{eqnarray} \label{v7}
\overline{F} \equiv \frac{\sum_i F[i]A[i]}{\sum_iA[i]}
\end{eqnarray}
has the universal importance of its  accurate ({"strong"}) counterpart in Eq.(\ref{w5}), 
\begin{eqnarray} \label{v7a}
\la F \ra \equiv\frac{ \sum_i F[i]p[i]}{\sum_ip[i]}, 
\end{eqnarray}
in Eqs.(\ref{w4}) and (\ref{w5}). In the next Section we will show that this cannot be the case, and for a good reason.
%%%%%%%%%%%%%%%%%%%%%%%%%%%%%%%%%%%%%%%%%%%%%%%%%%
\section{Weak values and the uncertainty principle}
As discussed in Sect.VIII, an accurately measured mean  value of the projector onto an intermediate state
gives the relative frequency with which the path created by the meter is travelled, and so serves to distinguish it from other real paths present. What would the result of such measurement mean 
in the most inaccurate limit,  where the path in question exists only as a virtual one? The uncertainty principle states that two virtual paths cannot be distinguished  by physical means inside the single real pathway they are combined into. 
\newline
There appears to be a logical rather than a mathematical difficulty. To address it, consider a purely classical conundrum.
Suppose there are two drops of water on a glass surface, and we put a grain of send in the drop number one. 
The drops move closer to each other, and coalesce. In which of the two drops is the sand grain now? If asked, one would certainly reject the question as meaningless, given the new circumstances. But if an answer is required under pain of death, the individual might reply "inside the water drop number $100$, for all I care". The answer to a meaningless question may be  {\it anything at all}.
%%%%%%%%%%%%%%%%%%%%%%%%%%%%%%%%%%%%%%%%%
\begin{figure}[ht]
\epsfig{file=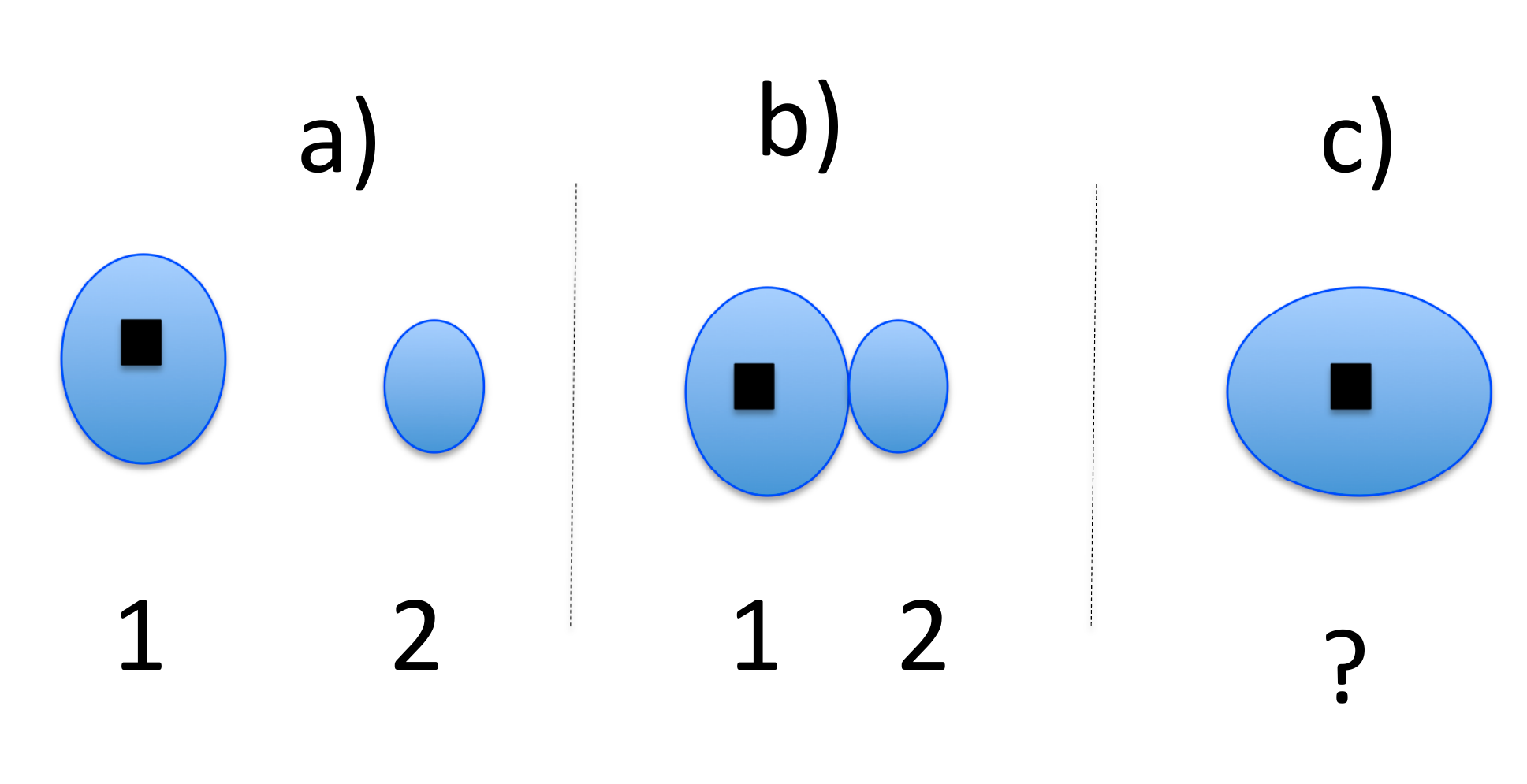, height=5cm, width= 14cm, angle=0}
\vspace{0pt}
\caption{To water drops on a smooth surface approach each other and coalesce. 
a) The grain of sand is in the first drop. b) The grain is still in the first drop. c) In which of the two drops is the grain now?}
\end{figure}
%%%%%%%%%%%%%%%%%%%%%%%%%%%%%%%%%%%%%%%%%
\newline
The above passage, illustrated in Fig.3,  is not entirely out of place in the present discussion. With two real paths created by an accurate meter,
in each trial  the pointer would point at zero or one, indicating the realisation of one of the two real possibilities.
If the meter is highly inaccurate, and the interference between two virtual paths is not destroyed, the pointer is equally likely to point at any number. If asked "was the first or the second virtual path travelled in this transition?", the meter, unable to politely decline to answer a meaningless question, points at, say, $100$. Importantly, this behaviour is a necessity, rather than a paradox.
It is the uncertainty principle that makes the question meaningless, and any other kind of answer would lead to a conflict with the principle.
\newline
Fully aware of the randomness  of an individual inaccurate result, the authors of \cite{Ah} focused on the meter's average reading  which, given the same initial an final states, is  a uniquely defined quantity, which should come up the same in every long series of trials. This is given by the real part  (or the  imaginary part, if a slightly different meter is used \cite{Ah},\cite{PLA16}) of what looks like an average (\ref{w5}), but obtained with a complex valued distribution, constructed from the path amplitudes $A[i]$.
We wish examine its meaning, bearing in mind the restrictions imposed by the UP.
\newline
Firstly, it should be remembered that Eq.(\ref{v5}) {\it expresses} the average $\overline{\xi}$ {\it defined} in a standard way by Eq.(\ref{x6}), in terms
of  terms of the systems amplitudes $\alpha[i]$. 
%The corresponding expression for the variance of $\xi$  .
Secondly, as discussed in Sect.VIII, mean pointer reading in an accurate measurement of a projector yields the relative frequency with which the real path in question is travelled. Thus with $N>>1$ trials performed, $\overline{\xi}N$ gives the number of trials in which the system has chosen the path. The result makes sense, since $p[i]\ge 0$, and  $\overline{\xi}N\le N$.
The same cannot be said about the amplitudes $A[i]$. Choosing $A[1]=1$ and $A[2]=-1.01$, and inserting them into Eq.(\ref{v7}),
we obtain $\overline{\xi}=-100$. An attempt to interpret $\overline{\xi}N$ as the number of cases where the first virtual paths is travelled in a series of $10^6$ trials gives a bizarre value of $-10^8$, another arbitrary answer to a meaningless question.
\newline
We need not limit ourselves to measuring projectors. 
Every functional taking the values $F[i]$ can be written as a sum $F[path]=\sum_jF[j] F^j[path]$, where $F^j[path]$ takes the values
$F^j[i]=\delta_{ij}$. In an accurate measurement, the mean value of $F^j[path]$ gives the relative frequency with which the $j$-th path is travelled. In an inaccurate measurement, its value will remain indeterminate, and with it also the value of $F[path]$.
\newline
Suppose a group of researchers measure the $z$-components of the spin of an electron,
$\sigma_z$,
at $t=T/2$, in a setup where the spin is pre- and post-selected at $t=0$ and $t=T$. Moreover, all researchers have different initial and final states $|\psi\ra$ and $|\phi\ra$. If the measurement is accurate, all researchers will be obtaining the results $-1$ and $1$, 
albeit in different proportions. They can all agree that the quantity they measure may take only these two values, and use this knowledge to develop theories of spin-1/2 particles, if needed.  If the measurement is weak, they can only agree that the spin they measure is able to take any real value at all.
\newline
Suppose further that the equipment is such that each researcher only has access to the average values, and not to the distributions. If the measurements are accurate, for all choices of $|\psi\ra$ and $|\phi\ra$ $\overline{\xi}$ would lie 
between $-1$ and $1$. This gives the knowledge of a useful fact that $\sigma_z $ represents a quantity whose value cannot be larger than one, or less than minus one, although  its discreet nature must be learned from some other source. If the measurement is weak, it can be shown \cite{PLA14}, that there is always a choice of $|\psi\ra$ and $|\phi\ra$ to give $\overline{\xi}$ any real value, positive, negative or zero. Thus, after trying all possible $|\psi\ra$ and $|\phi\ra$, the experimentators will remain in the state of maximal  ignorance regarding the properties of the measured quantity.
\newline
The mathematical trick which allows the UP to conceal the information about individual virtual paths is based on replacing the path probabilities with path amplitudes when calculating a weak value. It is not so much the fact that $A[i]$ are in general complex valued, 
but rather the absence of restrictions on the signs of their real and imaginary parts, that can cause the weak value of a projector to be $-100$ \cite{f3}. The mean and variance of a non-negative distribution give the estimates for the centre and the width of its region of support.  The mean and variance of a distribution which can change sign do not have this property, allowing $\overline{\xi}$ lie anywhere on the real axis, or obtain a zero variance even when the distribution is not a Dirac delta \cite{f3}. 
%%%%%%%%%%%%%%%%%%%%%%%%%%%%%%%%%%%%%%%%%%%%%%
\section{Weak values: over-interpretations and misconceptions}
Conclusions of the previous Section can be summarised as follows \cite{PLA16}.
A highly inaccurate "weak" measurement (WM) on a pre- and post-selected system is a perturbative scheme, in which an additional system (a vN meter) is weakly coupled to the observed one (e.g., a spin).
 The coupling is such that the mean value of one of the meter's variables (pointer position $\xi$) coincides with the real (or imaginary, if necessary \cite{Ah}, \cite{PLA16}) part
of a weighted sum of the relative amplitudes on the virtual paths connecting the system's initial and final states, 
\begin{eqnarray} \label{r1}
\overline{\xi} \sim \text{Re}(\text{Im}) \sum_{i}F[i]\al[i], \q \al[i]\equiv A[i]/ \sum_{i}A[i].
\end{eqnarray}
In Eq.(\ref{r1}) the paths $\{i\}$, $i=1,2...$ and the weighs $F[i]$ are determined by the nature of the measured quantities.
If is is possible (it may  not be) to measure directly a functional $F^j$, whose values are zero on all paths except the path number $j$, $F^j[i]=\delta_{ij}$
the mean reading of the weak pointer will yield precisely the real (imaginary) part of the amplitude $\al_j$. In short, weak measurements measure probability amplitudes.
%and, we argue, have no more significance than that.
\newline
A certain amount of confusion, which led  to a plethora of recent publications on this still fashionable subject (for a recent review see \cite{Dress} and Refs. therein), has its roots in Ref. \cite{Ah}
where the weak values, defined in \cite{Dress} as  {\it "...complex numbers that one can assign 
to the powers of a quantum observable operator  $\hat{A}$ using two states, an initial state $|i\ra$ called}  the preparation {\it or}  preselection, {\it and a final state $|f\ra$ called the} postselection",  were first introduced. 
The authors of \cite{Ah} considered a weak measurement of the $z$-component of a spin-1/2, similar to that described in Sect.V,
and concluded that {\it "...the usual measurement procedure for preselected and postselected  ensembles of quantum systems gives unusual results."} The title question of the Ref.\cite{Ah} "How the result of measurement of a component of the spin of a spin-1/2 particle can turn out to be 100?" has, however,  a simple answer. This would happen if the meter measures the difference of relative amplitudes $\al[1]-\al[2]$, for an improbable transition, for which $|A[1]+A[2]|^2=|\la \phi|\psi\ra|^2$ is small.
Surprise at the "unusual results" is also misplaced: the authors of \cite{Ah} might as well obtain any other value, large or small, negative, positive or zero. Indeed, the opposite would be surprising, as finding only the values between $-1$ and $1$ would contradict the uncertainty principle, as discussed in the previous Section.
\newline 
With weak values described only vaguely as a "new concept" \cite{Ah}, some authors proposed to use them as a tool for resolving 
"counterfactual paradoxes". One simple example is the so-called "three-box case"  \cite{AhBOOK}, \cite{Ah3b}, \cite{DS3b}, where the final state can be reached via three virtual pathways with amplitudes 
\begin{eqnarray} \label{r2}
A[1]=C, \q A[2]=-C, \q \text{and} \q A[3]=C,
\end{eqnarray}
where $C$ is a complex number.
An accurate measurement of a functional $F^1$, 
\begin{eqnarray} \label{r2a}
F^1[1]=1, \q F^1[2]=F^1[3]=0, 
\end{eqnarray}
always finds the system taking the first path.
On the other hand, an accurate measurement of a functional $F^3$, 
\begin{eqnarray} \label{r2b}
F^3[3]=1, \q F^3[1]=F^3[2]=0, 
\end{eqnarray}
always finds the system taking the third path, and the authors of \cite{AhBOOK}  alert the reader to {\it "...the peculiarity of having one particle in several places simultaneously even in the stronger sense than in the double slit experiment..."}. Admitting that the two results were obtained in different physical circumstances, the authors of \cite{AhHARDY} rely on the  WM {\it "... to test... the assertions otherwise regarded a counterfactual"}. Their reasoning is as follows.
Since WM perturb the system only slightly, it is possible to conduct several of them at the same time \cite{AhHARDY} ( or, we add, even 
perform full tomography of the studied transition \cite{PLA16}). Simultaneous WM of $F_1$ and $F_3$ by two meters yield
$\overline{\xi}_1=1$ and $\overline{\xi}_2=1$, and seemingly confirm the particle's presence in two places at the same time.
\newline
A closer look suggests a far simpler explanation of this "peculiarity". Different strong measurements "engineer" two completely different classical ensembles (cf. the passage at the end of Sect.VIII). This becomes clear if one adds all remaining final states, and the real pathways leading to them \cite{DS3b}. The fact that in two different setups the same final state is always reached via  paths of a different kind is not surprising at all. (We would not give it a second thought if the setups were built from the tubes and connectors of Sect.IV.)  
%\newline
So what is tested by the simultaneous WM of $F_1$ and $F_3$? We have seen that their results would be 
\begin{eqnarray} \label{r2c}
\overline{\xi}_1=Re\al[1]
%\frac{A[1]}{A[1]+A[2]+A[3]}
=1, \q \text {and}\q 
\overline{\xi}_2=Re\al[3]
%\frac{A[3]}{A[1]+A[2]+A[3]}
=1.
\end{eqnarray}
Since the relative amplitudes always add up to one, $\al[1]+\al[2]+\al[3]=1$, we also learn that $Re \al[2]=-1$. Far from asserting that the particle is in "two places", the weak values give us the values of the real parts of the relative amplitudes.
This information is not yet sufficient even for predicting the results of the two accurate measurements, since for this we also need the imaginary parts of $\al[i]$. Suppose that by making a WM of different kind we established that $Im \al[i]=0$, $i=1,2,3$. Now we know that all $\al$'s are real, and $\al[1]=-\al[2]=\al[3]$, and the standard rules of quantum mechanics tell us what would happen 
in two {\it mutually exclusive} situations, where the paths $\{2\}$ and $\{3\}$ are combined into a real pathway $\{2+3\}$, 
or  $\{1\}$ and $\{2\}$ are combined into $\{1+2\}$, as happens in the two accurate measurements mentioned above. 
The last shred of mystery, if there was one, disappears when we recall that the perturbation theory typically gives information about both the  moduli and the phases of amplitudes and wave function, and the WM scheme is the rule rather than the exception \cite{PLA16}.
\newline
Finally a  much criticised  \cite{COMM}, \cite{PLA14} paper entitled "How the result of a single coin toss can turn out to be 100 heads" \cite{CRAP}, the authors argue that { \it "weak values are not inherently quantum but rather a purely statistical feature of pre- and postselection with disturbance"}. Without going into the details of the author's reasoning, or into the arguments of their critics, 
one sees that the claim of \cite{CRAP} must be wrong. To obtain a mean value which lies outside the region of support of a distribution, the distribution must change sign \cite{f3}. Such distributions naturally appear in quantum mechanics, whose basic building blocks are complex valued probability amplitudes \cite{Feyn2}. They may not, however, appear in a classical theory where all physically meaningful averages are taken with non-negative probabilities.
\newline
The parallel between the titles of \cite{Ah}, and the Ref.\cite{CRAP}, which followed it twenty six years on, is not coincidental.
The fallacy of weak classical values in \cite{CRAP} stems naturally from the suggestion \cite{Ah} that the combinations of the transitions amplitudes, which arise when a measurement is extremely inaccurate, amount to a "new concept" of a quantum variable, capable of shedding light on a new physical reality.
%%%%%%%%%%%%%%%%%%%%%%%%%%%%%%%%%%%%%%%%%%%%%%%%%%%
\section{Conclusions}
In summary, we have shown that the  application of a sequence of quantum measurements creates a stochastic network.
A network is classical in the sense that the probabilities for all observable events have, of course, all the usual properties 
of probabilities in a classical theory. 
A network is quantum in its origin, and the action of a meter, or meters,  can be seen as a way of  "engineering" real scenarios, 
endowed with probabilities, from the virtual paths of the system, for which only probability amplitudes are available. 
A network is conveniently described in terms of  functionals, which take numerical values on the virtual paths, and whose values the measurements are set to determine. 
In the case of von Neumann and von Neumann-like meters considered here, 
the real pathways are determined by:

(i) the set of virtual paths available for the system with if no meter is present. 
%These paths are defined in the representation(s) of the measured quantity(ies), e.g., by expanding in the eigenstates of the operators 
%$\hat{A}$ and $\hat{A}$ 

(ii) the values of the measured functional. Two paths sharing the same value cannot be distinguished by the meter(s).

%Two paths sharing the same value (e.g., the paths $\{1\}$ and $\{4\}$ in the above example)
%are not distinguished by the measurement, and can be combined into a single pathway we will denote $\{1+4\}$.
(iii) the form and width of the initial state of the pointer, $G(\xi)$, which determine the accuracy of the measurement.

Highly accurate measurements turn all virtual paths, that can be distinguished, into real ones, travelled with the probabilities 
proportional to the squared moduli of the corresponding probability amplitudes.
If a functional whose value is $1$ on one of the virtual paths, and $0$ on all others, is measured accurately, its mean value 
yields the relative frequency with which the path is travelled. 

If the same functional is measured in a highly inaccurate manner, its mean (weak) value coincides with the real (or imaginary) part of the relative amplitude, corresponding to the selected path. Since there no {\it apriori} restrictions on the values of the relative amplitudes, it is always possible to find a transition in which the result of  a weak measurement would be any desired number, large, small, positive or negative. This is a necessary consequence of the uncertainty principle, and the "unusual" weak values lying outside the spectrum of the operator $\hat{A}$ in the example given in Sect. V, are by no means unusual. Many of exotic interpretations of the weak measurements results stem from the failure to identify the weak values with the path amplitudes, or their linear combinations, and are easily dismissed once the problem is reformulated in the language of standard quantum mechanics.
%%%%%%%%%%%%%%%%%%%%%%%%%%%%%%%%%%%%%%%
\section {Acknowledgements}  Support of the Project Grupos Consolidados
UPV/EHU del Gobierno Vasco (IT-472-10) and the MINECO
Grant No. FIS2015-67161-P, as well as useful discussions with Prof. E. Akhmatskaya are gratefully acknowledged.
%\end{document}
 

\begin{thebibliography}{10}
\bibitem{FOOTG} The choice of $G(\xi)$ as a real function ensures that both the mean momentum and the mean position of 
 the pointer are zero. In the classical limit, a von Neumann meter does not perturb the motion of the monitored system, provided its momentum is set exactly zero. Quantally, this condition can be met only approximately. The spread of the momenta around the zero mean accounts, therefore, for the inevitable perturbation the meter would incur. Making $G$ broad in the coordinate space would 
 minimise this perturbation, at the expense of the measurement's accuracy.
\end{thebibliography}
\end{document}